\documentclass[prb,twocolumn,groupedaddress,amsmath,floatfix]{revtex4}
\usepackage{graphicx}
\usepackage{bm}

\newcommand{\curH}{\hat{\cal H}}

\newcommand{\phdag}{{\phantom{\dagger}}}

\newcommand{\vol}{{V}}
\newcommand{\kf}{k_{\rm F}}

\newcommand{\ef}{\epsilon_{\rm F}}

\newcommand{\ch}{\hat{c}}

\newcommand{\bk}{{\bf k}}
\newcommand{\bq}{{\bf q}}
\newcommand{\bp}{{\bf p}}

\newcommand{\br}{{\bf r}}
\newcommand{\be}{\begin{equation}}
\newcommand{\ee}{\end{equation}}
\newcommand{\bea}{\begin{eqnarray}}
\newcommand{\eea}{\end{eqnarray}}
\newcommand{\bse}{\begin{subequations}}
\newcommand{\ese}{\end{subequations}}

\newcommand{\bQ}{{\bf Q}}

\newcommand{\sfm}{SF$_{\rm M}$\,}

\newcommand{\fermiint}{\lambda}

\newcommand{\as}{a_s}
\input{epsf}

\begin{document}
\title{Comment on \lq\lq Superfluid stability in the BEC-BCS
  crossover\rq\rq}
  
\author{Daniel E.~Sheehy and Leo Radzihovsky}
\affiliation{
Department of Physics, 
University of Colorado, 
Boulder, CO, 80309}
\date{August 7, 2006}
\begin{abstract}
  We point out an error in recent work by Pao, Wu, and Yip [Phys. Rev.
  B {\bf 73}, 132506 (2006)], that stems from their use of a necessary
  but not sufficient condition [positive compressibility (magnetic susceptibility) and superfluid
  stiffness] for the stability of the ground state of a polarized
  Fermi gas.
  As a result, for a range of detunings their proposed ground-state solution is a local maximum
  rather than a minimum of the ground state energy, which thereby
  invalidates their proposed phase diagram for resonantly interacting
  fermions under an imposed population difference.
\end{abstract}
\maketitle

There has been considerable recent interest in paired superfluidity of
fermionic atomic gases under an imposed spin
polarization~\cite{Zwierlein05,Partridge05}, i.e., when the numbers
$N_\uparrow$ and $N_\downarrow$ of the two atomic species undergoing
pairing are different.  Along with the detuning $\delta$ of the
Feshbach resonance (controlling the strength of the interatomic
attraction), the population difference $\Delta N = N_\uparrow -
N_\downarrow$ is an experimentally-adjustable \lq\lq knob\rq\rq\ that
allows the study of novel regimes of strongly-interacting fermions.

\begin{figure}
\epsfxsize=9.5cm
\centerline{\epsfbox{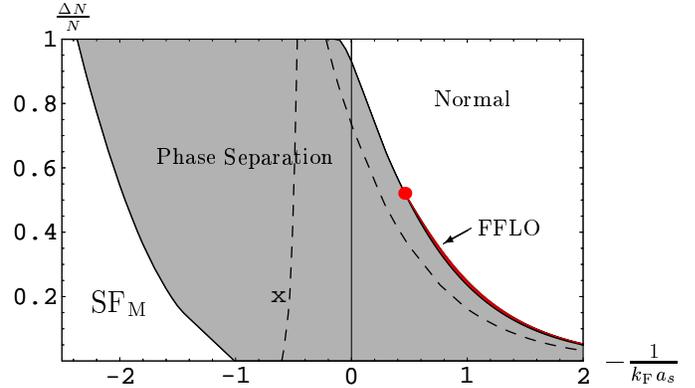}}
\vskip-.45cm
\caption{(Color online) Polarization $\Delta N/N$ vs. detuning $-\frac{1}{\kf \as}$ 
  phase diagram of the one-channel model (appropriate for width
  $\gamma\gg 1$) showing regimes of FFLO, superfluid (confined to the
  $\Delta N = 0$ axis), magnetized superfluid (\sfm), and phase
  separation. The dashed lines are the purported phase boundaries
  reported in Ref.~\onlinecite{Pao}.  As we show\cite{longpaper}, for
  the entire shaded region single-component (uniform) solutions to the
  stationarity conditions Eqs.~(\ref{eq:stationarity}) are {\it local
    maxima\/} of $E_G(\Delta)$ (and therefore unstable, leading to
  phase separation), shown for the point marked with the \lq\lq
  x\rq\rq\ in Fig.~\ref{fig:sarmaplot}.}
\label{fig:phasediagram}
\end{figure}

A crucial question concerns the phase diagram of resonantly
interacting fermions as a function of $\delta$ and imposed $\Delta N$.
Two early theoretical studies that have addressed this issue are the work 
by Pao, Wu, and Yip~\cite{Pao} on the one-channel model of interacting fermions and 
our work~\cite{shortpaper} on the two-channel model of interacting fermions,
with the details and extensions of the latter presented in our recent
preprint.~\cite{longpaper}
Apart from the Fulde-Ferrell-Larkin-Ovchinnikov (FFLO) phase appearing
over a thin sliver of the phase diagram in
Ref.~\onlinecite{shortpaper} (explicitly not considered by Pao, Wu, and Yip),
for the positive-detuning BCS and crossover regimes the phase diagrams
in these manuscripts are in qualitative agreement~\cite{although}.
However, at negative detuning in the BEC regime, the phase diagrams
are qualitatively different.  In particular, in Fig.~4 of
Ref.~\onlinecite{Pao}, stable superfluidity is claimed to exist {\it
  above\/} and to the left of a {\it nearly vertical\/} and positively
sloped phase boundary that we re-plot here in
Fig.~\ref{fig:phasediagram} as a dashed line at negative detuning.  In
qualitative contrast, in our work, Ref.~\onlinecite{shortpaper}, we
found that stable superfluidity exists only {\it below\/} and to the
left of a negatively-sloped phase boundary (see
Fig.~\ref{fig:phasediagram}).

What is the source of this {\em qualitative} discrepancy? Although
Ref.~\onlinecite{Pao} and Ref.~\onlinecite{shortpaper} use different
models of resonantly-interacting fermions, the close relationship
between the one- and two-channel models (particularly within the
mean-field approximation) implies that they should yield qualitatively
similar phase diagrams.  Furthermore, we have extended our original
two-channel model study\cite{shortpaper} to that of a
one-channel model\cite{longpaper} and, as expected, found results in
qualitative agreement with those in our Letter\cite{shortpaper}, but
in disagreement with that of Ref.~\onlinecite{Pao}.  

Indeed, as we explicitly show here, the origin of the discrepancy is
that the criterion for stability of the superfluid phase used in
Ref.~\onlinecite{Pao} (based on positivity of magnetic
susceptibilities) is a necessary but {\em not} sufficient condition
for stability and does not ensure~\cite{unfortunately,Iskin,Lamacraft} that the
state is even a local minimum of the energy. 
This thereby leads to
incorrect phase boundaries in both the BEC and BCS regimes, although
below we shall focus on the phase boundary inside the BEC regime,
where the error is qualitative and most pronounced.

Thus, much of what is claimed to be a \lq\lq stable superfluid\rq\rq\ 
in Ref.~\onlinecite{Pao} (in the negative-detuning BEC regime) is
actually unstable to phase separation. This is illustrated in the
correct $T=0$ mean-field phase diagram, Fig.~\ref{fig:phasediagram},
for the one-channel model (quantitatively consistent with other recent
work~\cite{Gu,Parish} and derived in detail in
Ref.~\onlinecite{longpaper}), plotted as a function of the
dimensionless parameter $-(\kf \as)^{-1}$ (proportional to the
Feshbach resonance detuning $\delta$, with $\kf$ the Fermi wavevector
and $\as$ the s-wave scattering length) and the polarization $\Delta
N/N = (N_\uparrow - N_\downarrow)/(N_\uparrow + N_\downarrow)$.

Our starting point is the single-channel model Hamiltonian (studied in
Ref.~\onlinecite{Pao}) for two resonantly-interacting species of
fermion $\ch_{\bk\sigma}^\phdag$ ($\sigma = \uparrow,\downarrow$):
\be
\curH = \sum_{\bk,\sigma}\epsilon_k\ch_{\bk\sigma}^\dagger  \ch_{\bk\sigma}^\phdag 
+ \frac{\fermiint}{\vol}\sum_{\bk\bq\bp} \ch_{\bk\uparrow}^\dagger \ch_{\bp\downarrow}^\dagger 
\ch_{\bk+\bq\downarrow}^\phdag
\ch_{\bp-\bq\uparrow}^\phdag,
\label{eq:singlechannelintro}
\ee
where $\epsilon_k = k^2/2m$, $m$ is the fermion mass, and $\vol$ is
the system volume.  A broad Feshbach resonance is modeled by an
attractive interaction $\fermiint<0$, the magnitude of which increases
with decreasing Feshbach resonance detuning.

The equilibrium ground state of a many particle system in the
grand-canonical ensemble (with chemical potentials $\mu_\uparrow$ and
$\mu_\downarrow$) at $T=0$ is characterized by the grand thermodynamic
potential $\Omega(\mu_\uparrow,\mu_\downarrow)$ defined by~\cite{Forbes}
\bea
\label{eq:omega}
&&\Omega(\mu_\uparrow,\mu_\downarrow) = 
{\rm min}\left[\langle\hat{H}\rangle\right],
\eea
with the grand-canonical Hamiltonian 
\bea
&&\hat{H} \equiv \curH - \mu_\uparrow \hat{N}_\uparrow - \mu_\downarrow \hat{N}_\downarrow.
\label{eq:curH}
\eea
Here, $\hat{N}_\sigma \equiv \sum_\bk \ch_{\bk\sigma}^\dagger
\ch_{\bk\sigma}^\phdag $ is the number operator for fermion species
$\sigma$ and the minimization in Eq.~(\ref{eq:omega}) is over all
possible ground states.

The standard (BCS-type) mean-field approximation that we shall utilize
(as done in Ref.~\onlinecite{Pao}) amounts to assuming a restricted
class of possible many-body ground-states self-consistently
parametrized by the ground-state expectation value
\be
\Delta = \lambda \langle \ch_{\downarrow}^\phdag(\br) \ch_{\uparrow}^\phdag(\br) \rangle.
\label{deltaexp} 
\ee
For pairing amplitude $\Delta \neq 0$, this class includes both 
weakly-paired BCS-type and strongly-paired molecular BEC-type pairing
order. For $\Delta = 0$ it is an unpaired Fermi gas.

Once this mean-field approximation has been made, it is
straightforward~\cite{shortpaper,longpaper} to compute the variational
ground-state energy, which is the expectation value $E_G(\Delta) =
\langle \hat{H} \rangle$.  Evaluating this expectation value, and
converting momentum sums to integrals, we find
\bea
&&E_G(\Delta)/\vol = -\frac{m}{4\pi \as\hbar^2}\Delta^2 + 
\int \frac{d^3k}{(2\pi\hbar)^3} ( \xi_k - E_k +\frac{\Delta^2}{2\epsilon_k}) 
\nonumber 
\\
&& \qquad  + \int \frac{d^3k}{(2\pi\hbar)^3} (E_k-h) \Theta(h-E_k),
\label{eq:havgBCSp}
\eea
where $\xi_k \equiv \epsilon_k -\mu$, $E_k \equiv
\sqrt{\xi_k^2+\Delta^2}$, and we have defined the chemical potential
$\mu = \frac{1}{2}(\mu_\uparrow + \mu_\downarrow)$ and the chemical
potential difference $h=  \frac{1}{2}(\mu_\uparrow - \mu_\downarrow)$.  We have also
exchanged the bare interaction parameter $\lambda$ for the vacuum
s-wave scattering length $\as$ given by
\be
\frac{m}{4\pi \as\hbar^2} = \frac{1}{\lambda}+\int \frac{d^3 k}{(2\pi\hbar)^3} \frac{1}{2\epsilon_k}.  
\ee

The determination of the mean-field phase diagram using
Eq.~(\ref{eq:havgBCSp}) is conceptually quite simple.  According to
Eq.~(\ref{eq:omega}), the ground state at a particular $\mu_\uparrow$
and $\mu_\downarrow$ (or, equivalently, $\mu$ and $h$) is given by the
minimization of $E_G(\Delta)$ with respect to the pairing amplitude
$\Delta$ that can be taken to be real.  Any such minima of course
satisfy the stationarity constraint or gap equation
[equivalent to Eq.~(\ref{deltaexp})]
\bse
\label{eq:stationarity}
\be
\label{eq:gap}
0 = \frac{\partial E_G}{\partial \Delta},
\ee
where we emphasize that the derivative is taken at fixed $\mu$ and $h$.  Since experiments are
conducted at fixed atom number, we must augment Eq.~(\ref{eq:gap})
with the number constraint equations $N_\sigma = \langle
\hat{N}_\sigma \rangle$.  Examining Eq.~(\ref{eq:curH}), we see that
the constraints can be rewritten as
\bea
\label{eq:num}
N&=&-\frac{\partial E_G}{\partial\mu},
\\
\label{eq:pol}
\Delta N&=&-\frac{\partial E_G}{\partial h},
\eea
\ese
with the total particle number $N = N_\uparrow + N_\downarrow$ and
population difference $\Delta N= N_\uparrow - N_\downarrow$.

Our key point (apparently missed by the authors of Ref.~\onlinecite{Pao})
is that not every simultaneous solution of the gap and number equations,
Eqs.~(\ref{eq:stationarity}), corresponds to a physical ground state of
the system; the additional criterion is that the solution $\Delta$
must also be a {\em minimum} of $E_G(\Delta)$ at fixed $\mu_\sigma$.
The verification that an extremum solution is indeed a minimum is
particularly essential when there is the possibility of a first-order
transition, with $E_G(\Delta)$ exhibiting a local maximum that separates
local minima, as is the case for a polarized Fermi gas, studied here
and in Refs.~\onlinecite{Pao,shortpaper,longpaper}. 

Analyzing Eqs.~(\ref{eq:stationarity}), we find that for sufficiently large
$\Delta N$ in the positive-detuning BCS regime and for $\Delta N = N$ 
in the BEC regime, a solution to Eqs.~(\ref{eq:stationarity}) may be found that
minimizes $E_G(\Delta)$ at $\Delta =0$, indicating a polarized normal
phase.  Also, for sufficiently low detuning in the BEC regime, a
polarized molecular superfluid (\sfm) solution exists that minimizes
$E_G(\Delta)$ at $\Delta \neq 0$.  A more general
analysis~\cite{shortpaper,longpaper,noteq} shows that a
periodically-paired FFLO solution is the ground state
over a thin range of polarization values at sufficiently large
positive detuning.

However, over the large shaded portion of the phase diagram, at
intermediate detuning and polarization, we find that it is not
possible to satisfy Eqs.~(\ref{eq:stationarity}) with a (homogeneous,
single component) minimum of $E_G(\Delta)$.
For the corresponding range of parameters the system
phase-separates~\cite{Bedaque} into two coexisting ground states (that
are degenerate minima of $E_G[\Delta]$).  The resulting
phase-separated state can be explicitly accounted for by generalizing
the ground-state ansatz to include the possibility of such an
inhomogeneous mixture~\cite{longpaper}.
%

\begin{figure}
\epsfxsize=9.5cm
\centerline{\epsfbox{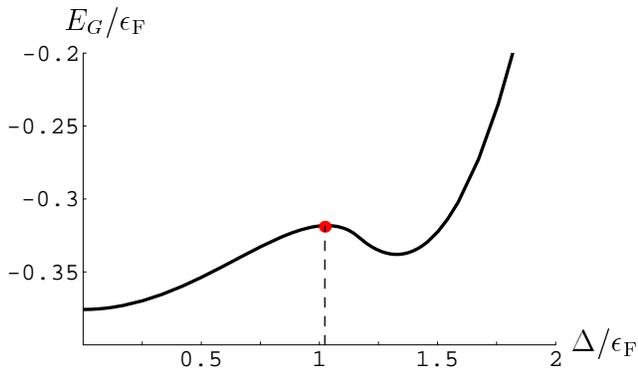}}
\vskip-.75cm
\caption{Plot of $E_G(\Delta)$ at fixed $\mu = -0.013\ef$ and $h = 1.16\ef$, 
  normalized to the Fermi energy $\ef$, for coupling $g = \frac{1}{\kf
    \as} = 0.63$, so that Eqs.~(\ref{eq:stationarity}) yield a
  stationary solution with $\Delta N/N = 0.2$ and $\Delta =1.02$.  As
  seen in Fig.~\ref{fig:sarmaplot}, however, this solution is a {\it
    local maximum\/} (saddle point) of $E_G(\Delta)$.  At this point
  ($\frac{1}{\kf \as} = 0.63$ and $\Delta N/N = 0.2$) in the phase
  diagram (illustrated by an ``x'' in Fig.\ref{fig:phasediagram}) the
  actual mean-field ground state is a phase-separated mixture of a
  superfluid and a normal state.}
\label{fig:sarmaplot}
\end{figure}

The contrasting strategy of Pao, Wu, and Yip~\cite{Pao} is to find solutions
of Eqs.~(\ref{eq:stationarity}) for all values of $N$, $\Delta N$, and
$-\frac{1}{\kf \as}$, some of which do not correspond to ground
states.  The unphysical (unstable) solutions are then discarded
(thereby determining the phase boundaries plotted in
Ref.~\onlinecite{Pao}) based on criteria of the positivity of magnetic
susceptibilities (atomic compressibilities) and the superfluid
stiffness. However, as we now discuss, these stability criteria are
necessary but {\em not sufficient} (i.e., not generally restrictive
enough) to ensure that a solution to Eqs.~(\ref{eq:stationarity}) is
indeed a minimum of $E_G(\Delta)$. In contrast, it can be shown that
the converse is true, namely that the minimization of $E_G(\Delta)$
ensures that the local susceptibilities are positive
definite.\cite{susceptibilitynote}

The stability criteria~\cite{sfstiffnessnote} used by Pao, Wu, and Yip can be
understood by examining Fig.~3 of Ref.~\onlinecite{Pao}.  The solid
lines in this figure correspond to solutions of
Eqs.~(\ref{eq:stationarity}) at different values of the coupling $g
\equiv \frac{1}{\kf \as}$.  In particular they plot $h/\ef$ (with
$\ef$ the Fermi energy, related to the density $n = N/V$ by $n =
\frac{4}{3} c\ef^{3/2}$ with $c = m^{3/2}/\sqrt{2}\pi^2\hbar^3$) as a
function of the polarization $n_d/n = \Delta N/N$ (at fixed density),
where $n_d = \Delta N/\vol$ is the magnetization.  At positive and
intermediate detunings (the bottom curves of Fig.~3 of Pao, Wu, and Yip),
they find solutions satisfying $\frac{\partial h}{\partial n_d}\big|_n <0$
and correctly conclude that such solutions (having a negative magnetic
susceptibility) are unstable.  However, at sufficiently negative
detuning ($g\agt0.5$) Pao, Wu, and Yip.  find solutions to
Eqs.~(\ref{eq:stationarity}) with a positive susceptibility
$\frac{\partial h}{\partial n_d}\big|_n >0$, and based on their criterion
(erroneously) conclude that these solutions indicate a stable magnetic
superfluid ground state.  They then define a phase boundary in the BEC
regime (the leftmost dashed curve of Fig.~\ref{fig:phasediagram}), to
a stable magnetized superfluid, by where $\frac{\partial h}{\partial
  n_d}\big|_n$ changes sign.

However, our explicit calculation of $E_G(\Delta)$ (plotted in
Fig.~\ref{fig:sarmaplot}) for one such solution (with $\frac{1}{\kf
  \as} = g = 0.63$ and $n_d/n = 0.2$, indicated with an ``x'' in
Fig.~\ref{fig:phasediagram} and corresponding to a point on the
uppermost solid curve of Fig.~3 of Ref.~\onlinecite{Pao}), purported
by Pao, Wu, and Yip to be stable (to the left of their proposed stability
boundary), shows that in fact this solution (indicated with a  dot
in Fig.~\ref{fig:sarmaplot}) is a {\it local maximum} and therefore
does {\em not} represent a ground state.  This solution was obtained
by numerically solving Eqs.~(\ref{eq:stationarity}) at $g = 0.63$ and
$\Delta N/N =0.2$, yielding $\mu =-0.013\ef$, $h = 1.16 \ef$ and
$\Delta = 1.02 \ef$, the latter two values consistent with Figs.~2 and
~3 of Pao, Wu, and Yip, showing that we are indeed reproducing a solution
claimed to be stable by Pao, Wu, and Yip. 
 Thus, the method used by Pao, Wu, and Yip has not correctly located the global 
minimum of the ground-state energy; indeed, it has not even found a local
minimum.   

Although it might appear from the plot of $E_G(\Delta,\mu,h)$
(Fig.~\ref{fig:sarmaplot}) that the true ground state is an unpaired
($\Delta=0$) normal state, this state does not satisfy
Eqs.~(\ref{eq:num}) and (\ref{eq:pol}); thus, it is also not the
ground state at this coupling and polarization.  Indeed, as noted
above, we find that it is impossible to minimize $E_G(\Delta)$ while
satisfying Eqs.~(\ref{eq:num}) and (\ref{eq:pol}) at this coupling and
polarization (marked by an ``x'' in Fig.~\ref{fig:phasediagram}), nor
anywhere inside the shaded region in the phase diagram
Fig.~\ref{fig:phasediagram}, indicating the absence of a uniform
solution.
The true mean-field ground state everywhere in the shaded region is a
phase-separated mixture of two phases of different densities in
chemical equilibrium such that the total number and polarization
constraints are satisfied~\cite{shortpaper,longpaper}.

We note that a ground state determined by minimizing $E_G(\Delta)$ at
a particular $\mu_\uparrow$ and $\mu_\downarrow$ automatically
satisfies the condition of having a positive magnetic susceptibility
(compressibility).  Indeed, it is straightforward to generally
show~\cite{Sewell,susceptibilitynote} that, since
$\Omega(\mu_\uparrow,\mu_\downarrow)$ is concave
downwards~\cite{Forbes}, the eigenvalues of the susceptibility matrix
\be
\chi \equiv \begin{pmatrix}
\frac{\partial{N_\uparrow}}{\partial{\mu_\uparrow}}
& \frac{\partial{N_\uparrow}}{\partial{\mu_\downarrow}}
\cr \frac{\partial{N_\downarrow}}{\partial{\mu_\uparrow}}
&\frac{\partial{N_\downarrow}}{\partial{\mu_\downarrow}}
\end{pmatrix} = - \begin{pmatrix}
\frac{\partial^2 \Omega}{\partial \mu_\uparrow^2}
& \frac{\partial^2 \Omega}{\partial \mu_\uparrow\partial \mu_\downarrow}
\cr\frac{\partial^2 \Omega}{\partial \mu_\uparrow\partial \mu_\downarrow}
&\frac{\partial^2 \Omega}{\partial \mu_\downarrow^2}
\end{pmatrix},
\label{eq:condition}
\ee
are positive in the ground state.  An equivalent stability criterion was 
derived in Ref.~\onlinecite{Viverit} by considering stability against
local density variations.  
The procedure used by Pao, Wu, and Yip, however, did not amount to analyzing Eq.~(\ref{eq:condition});
as discussed above, the phase diagram of Pao, Wu, and Yip
 was obtained by computing 
$\frac{\partial h}{\partial n_d}\big|_n$ which is {\it not\/} an equivalent condition.  If 
the authors had instead studied $\frac{\partial n_d}{\partial h}\big|_\mu$, they would have 
found that the solution plotted in Fig.~\ref{fig:sarmaplot} has a negative magnetic susceptibility
in the grand-canonical ensemble (and therefore is unstable). 
However, we must emphasize the important point that {\it any\/}  
particular extremum solution may have a positive magnetic susceptibility and still
not be the ground state. The simplest example
of this is the normal Fermi gas state ($\Delta =0$), which satisfies
the gap  and number-constraint equations {\it everywhere\/} in the phase diagram (including
$\Delta N=0$) and has a positive magnetic susceptibility, but is only
the actual ground state (a minimum of $E_G(\Delta)$) at sufficiently large
$\Delta N$.  
Thus, if the authors of Ref.~\onlinecite{Pao} had computed the eigenvalues of Eq.~(\ref{eq:condition}) 
instead of $\frac{\partial h}{\partial n_d}\big|_n$, they would have been able to discard some
of the erroneous solutions plotted in Fig.~3 of Ref.~\onlinecite{Pao}.  However, {\it in general\/}, 
Eq.~(\ref{eq:condition}) is still not sufficient and the most correct scheme is to use
Eq.~(\ref{eq:omega}), i.e., to find the global minimum, in the grand-canonical ensemble, of
the mean-field ground-state energy.

We conclude by noting that, although a mean-field analysis of the
one-channel model is not expected to be quantitatively accurate near
the resonance position where $\kf |\as| \to \infty$, it is expected to
yield a qualitatively correct description of a polarized
resonantly-interacting Fermi gas.  To summarize, we have shown that
while for equal species number ($\Delta N=h=0$) such analysis can
simply proceed by solving the gap and number equations
[Eqs.~(\ref{eq:stationarity})], the existence of first-order
transitions at $h\neq 0$ implies that the ground-state energy
$E_G(\Delta)$ exhibits {\it local maxima\/} as a function of $\Delta$,
yielding solutions to Eq.~(\ref{eq:gap}) that do not represent
physical ground states.

\smallskip
\noindent
{\it Acknowledgments\/} ---  
We gratefully acknowledge discussions with V. Gurarie and M. Veillette
as well as support from NSF DMR-0321848 and the Packard Foundation.


\begin{thebibliography}{10}
\bibitem{although} Although in this Comment we have focused on the BEC
  regime where the failure of the stability criteria used by Pao, Wu, 
and Yip\cite{Pao} is most drastic, the inadequacy is also obvious
  (although less drastic) on the BCS side, yielding a quantitatively
  incorrect phase boundary, as illustrated by dashed curves in
  Fig.~\ref{fig:phasediagram}.
\bibitem{unfortunately} Unfortunately, following
  Ref.~\onlinecite{Pao}, this erroneous (insufficient for stability)
  criterion was also more recently used by other authors yielding a
  similarly incorrect phase diagram; see for example
  Ref.~\onlinecite{Iskin}.  As also
discussed in Ref.~\onlinecite{Lamacraft}, such authors incorrectly identify spinodals
(where local stability of a metastable solution is lost) as being
phase boundaries.  Such an error yields, e.g., the rightmost
dashed-line boundary in Fig.~\ref{fig:phasediagram}.
\bibitem{noteq} 
  In Refs.~\onlinecite{shortpaper,longpaper}, we use a more general
  variational ansatz that is equivalent to assuming $\lambda\langle
  \ch_\downarrow^\phdag(\br)\ch_\uparrow^\phdag(\br)\rangle =
  \Delta_\bQ {\rm e}^{i\bQ\cdot \br}$, allowing the possibility of a
  periodically-modulated FFLO-type ground state.  Since the more
  general $\bQ\neq 0$ analysis, presented in
  Ref.~\onlinecite{longpaper} for the one-channel model, only yields
  stable FFLO states for a very thin window of parameters at positive
  detuning (indicated in red in Fig.~\ref{fig:phasediagram}), for
  simplicity (and because this assumption was also made by
 Pao, Wu, and Yip~\cite{Pao}) here we assume $\bQ =0$ at the outset.
\bibitem{susceptibilitynote} The concavity of
  $\Omega(\mu_\uparrow,\mu_\downarrow)$ [Eq.~(\ref{eq:omega})],
  implying positive eigenvalues of the susceptibility matrix
  Eq.~(\ref{eq:condition}) in the ground-state, is exactly true
  following general arguments~\cite{Forbes,Sewell} and is also true
  within the mean-field approximation.  To show the latter, we note
  that, within mean-field theory,
  $\Omega(\mu_\uparrow,\mu_\downarrow)$ is a minimization with respect
  to $\Delta$ of $E_G[\Delta,\mu_\uparrow,\mu_\downarrow]$, which, in
  turn, is a concave function (at fixed $\Delta$) of $\mu_\uparrow$
  and $\mu_\downarrow$ since it is the exact ground-state energy for
  the mean-field Hamiltonian.  The fact that the mean-field
  $\Omega(\mu_\uparrow,\mu_\downarrow)$ is a minimization of the
  concave function $E_G[\Delta,\mu_\uparrow,\mu_\downarrow]$ with
  respect to $\Delta$ then implies it is also concave.  To illustrate
  this, consider for simplicity setting $h=0$ so that the
  thermodynamic potential $\Omega(\mu)$ is a function only of $\mu$.
  Then one can show that the number susceptibility $dN/d\mu = -d^2
  E_G[\Delta_0(\mu),\mu]/d\mu^2$ satisfies [with $\Delta_0(\mu)$ the
  minimum of $E_G[\Delta,\mu]$]
%
\be
\frac{dN}{d\mu}  = -\frac{\partial^2 E_G}{\partial\mu^2} 
 +\frac{\partial^2 E_G}{\partial\Delta^2} \Big(
\frac{\partial \Delta}{\partial \mu}
\Big)^2\big|_{\Delta = \Delta_0}.
\nonumber
\ee
%
The second term of this equation is clearly positive at the
minimum, and the first term is positive by virtue of the above
concavity argument, guaranteeing that $dN/d\mu>0$ (positive
compressibility) in the ground state.  However, if $\Delta_0$ is only
a stationary point (e.g., a local maximum), this equation still
holds and shows that $dN/d\mu$ can be either positive or negative
depending on the relative magnitude of the two terms.  This change in
sign of $dN/d\mu$ (as takes place, for $dn_d/dh$, across the dashed curve in
Fig.\ref{fig:phasediagram}) must not be interpreted as a transition to a
stable superfluid phase as was done by Pao, Wu, and Yip\cite{Pao}.
\bibitem{sfstiffnessnote} Pao, Wu, and Yip (Ref.~\cite{Pao}) also consider a
  criterion based on the positivity of the superfluid stiffness but
  find this criterion to be even less stringent than the positive
  magnetic susceptibility criterion that we focus on here.
\end{thebibliography}
\end{document}